\documentclass[letterpaper, aps, tightenlines, nofootinbib, 11pt, preprintnumbers]{revtex4}
\pdfoutput = 1

\usepackage{setspace}
\usepackage[pdftex, bookmarks = false, colorlinks = true, linkcolor = blue, citecolor = purple]{hyperref}
\usepackage{shortcuts}
\usepackage{titlesec}

\usepackage[margin=2.5cm]{geometry}
\def\thesection{\arabic{section}}
\def\thesubsection{\arabic{subsection}}
\titleformat{\section}
  {\normalfont\Large\bfseries}{\thesection}{1em}{}
\titleformat{\subsection}
  {\normalfont\large\bfseries}{\thesection.\thesubsection}{1em}{}
\titleformat{\subsubsection}
  {\normalfont\it}{\thesubsection.\thesubsubsection}{1em}{}

\begin{document}
\begin{center}

\thispagestyle{empty}

\vspace*{5em}

{ \LARGE {\bf Singlet Vector Models on Lens Spaces }}

\vspace{1cm}

{\large\DJ or\dj e Radi\v cevi\' c}
\vspace{1em}

{\it Stanford Institute for Theoretical Physics and Department of Physics\\ Stanford University \\
Stanford, CA 94305-4060, USA}\\
\vspace{1em}
\texttt{djordje@stanford.edu}\\
\end{center}

\vspace{0.08\textheight}
\begin{abstract}
We present exact computations of partition functions of singlet vector models (infinite level Chern-Simons-matter theories) on lens spaces $L(p, 1)$. We identify light topological configurations and their spectra, and we comment on the relevance of our results in studying both the UV completions of Vasiliev's higher-spin theories and the dS/CFT correspondence in the large $N$ limit.
\end{abstract}

\preprint{SU-ITP-12/27}

\maketitle

\pagebreak
\setcounter{page}{1}

\section{Introduction}

The spectrum of singlet operators of a three-dimensional $N$-vector model corresponds to the spectrum of higher spin particles in four dimensions \cite{Klebanov:2002ja}. This observation suggests that (large $N$, zero 't Hooft coupling) conformal points of singlet vector models, defined more precisely below, possess gauge/gravity duals given by Vasiliev theories of classical higher spin gauge fields \cite{Klebanov:2002ja, Giombi:2009wh, Giombi:2012he, Vasiliev:1992av, Vasiliev:1995dn, Vasiliev:1999ba}. Neither the UV completion of Vasiliev theory nor its tentative dual, the strongly coupled limit of the singlet vector model, are fully understood. Recent work has shown that singlet vector models exhibit thermal phase transitions \cite{Shenker:2011zf} and possess light topological states \cite{Banerjee:2012gh} that are not present in the original Vasiliev theory. These results are data that can be used in order to understand the stringy embedding of Vasiliev theory. In this paper we add to these data a 
detailed study of topological configurations in the singlet vector model on lens spaces $L(p,1) = S^3/\Z_p$.

We use the phrase ``singlet vector model'' to refer to a theory of $N$ complex scalar fields whose spectrum is restricted to operators invariant under $U(N)$ rotations in the target space. Variations on this theme can be considered, but the analysis will be the same. A singlet vector model on a three-dimensional space with a compact spatial manifold can be given a local Lagrangian formulation in terms of a $U(N)$ Chern-Simons (CS) theory at infinite level coupled to fundamental matter \cite{Shenker:2011zf, Sundborg:1999ue, Aharony:2003sx}. The non-singlet states are projected out by Gauss' law, and the infinite level --- i.e.~the zero 't Hooft coupling --- is intended to make sure that the fluctuations in the gauge field do not couple to matter and that the theory remains conformal.  If we desire to formulate a duality between higher-spin gravity and a \emph{local} quantum field theory, then it seems that it is this infinite-level CS-matter theory that should be regarded as the dual to the Vasiliev theory in the bulk --- not just the matter sector. This is an important distinction, as putting a gauge theory on a topologically nontrivial space, e.g.~$S^1\times S^2$ or $\R \times \b T^2$, will give rise to nontrivial flat connections that may impact even the zero-coupling regime \cite{Shenker:2011zf, Banerjee:2012gh, Horowitz:2001uh}.

In this short note we analytically compute the free energy of the singlet vector model on the family of lens spaces $L(p, 1)$. As expected, we find a nontrivial structure of light, topologically induced configurations or ``states,'' as we will call them for convenience (even though these are \emph{not} the usual quantum states, since the lens space lacks a time direction). In particular, a $p^N$-like abundance of such states is found to always appear at very small  volume (large $p$) lens spaces. The calculation of the free energy on this family of spacetimes also allows us to comment on a na\"ive but inconsistent extrapolation of the dS/CFT correspondence \cite{Anninos:2011ui, Anninos:2012qw, Anninos:2012ft} to nontrivial topologies.

\section{Weakly coupled CS-matter theory on a lens space}

A lens space is a smooth three-manifold obtained by quotienting the sphere $S^3 = \{(z_1, z_2) \in \C^2\ \big|\ |z_1^2| + |z_2^2| = 1 \}$ by the freely acting, $\Z_p$-isomorphic group whose single generator can be taken to act as $(z_1, z_2) \mapsto (z_1 e^{2\pi i/p}, z_2 e^{-2\pi i/p})$. If we let $z_i = r_i e^{i\varphi_i}$, we may parametrize the sphere with Euler angles $(\theta, \phi, \psi)$ given by $\phi = \varphi_1 + \varphi_2$, $\psi = \varphi_1 - \varphi_2$ and $r_2/r_1 = \tan (\theta/2 + \pi/4)$. The geometry of $S^3$ in these coordinates can then be expressed as
\bel{
  \d s^2 = \frac14 \left(\d \theta^2 + \d \phi^2 + \d\psi^2 + 2\sin\theta \d\phi\d\psi \right),\quad \theta \in \left[-\frac\pi2, \frac\pi2\right],\ \phi \sim \phi + 2\pi,\ \psi \sim \psi + 4\pi.
}
The definition of the $\Z_p$ action implies that the lens space $L(p, 1) = S^3/\Z_p$ is the same manifold as above, except with $\psi \sim \psi + 4\pi/p$. In the rest of the paper we will refer to the quotienting procedure as ``orbifolding,'' and the $\psi$ circle will be the ``orbifolded circle.'' This is convenient nomenclature, but it should be stressed that it is slightly misleading:  there are no orbifold singularities on a lens space. 

We wish to study CS-matter theories on $L(p, 1)$ in the limit of infinite level. Similar gauge theories have already been studied in different contexts \cite{Hikida:2006qb, Aganagic:2002wv, Marino:2009dp, Marino:2011nm, Alday:2012au}. The infinite level breaks the path integral over gauge fields into a sum of fluctuations around flat connections (``vacua''). The vacua correspond to nontrivial holonomies on $L(p, 1)$, which are in turn indexed by the set of homomorphisms from $\pi_1(L(p,1)) = \Z_p$ to $U(N)$, modulo conjugations. Thus, each flat connection corresponds to a Wilson loop operator $V = \exp\left\{i\oint A_\psi \d\psi \right\}$ along the orbifolded circle. These Wilson loops inherit the group structure of $\Z_p$ and hence they must satisfy $V^p = 1$; moreover, quotienting by conjugations allows us to focus only on diagonal matrices $V$. Therefore, we may identify the set of flat connections (i.e.~the moduli space of vacua) with the set of diagonal matrices specified by a $p$-component vector $\b N 
= \{N_I\}$,
\bel{
  V =  \bigoplus_{I = 0}^{p - 1} e^{2\pi i I/p}\ \1_{N_I} = \trm{diag}\left(\underbrace{1,\ldots, 1}_{N_0}, \ \ldots,\ \underbrace{e^{2\pi i(p - 1)/p}, \ldots, e^{2\pi i(p - 1)/p}}_{N_{p - 1}}\right).
}
Henceforth the indices $I$, $J$, etc.~will always run from $0$ to $p - 1$. The set of vacua is thus indexed by all the partitions of $N$ into $p$ summands. Within each vacuum the gauge group is broken down to $\prod_I U(N_I)$, and the $N$-vector model splits into a direct sum of $N_I$-vector models, each having different boundary conditions along the orbifolded circle due to the existence of a nontrivial Berry phase. The discreteness of the set of vacua implies that, up to the just described sensitivity on the boundary conditions, matter completely decouples from the gauge fields at infinite level. 

The partition function of the singlet vector model on $L(p, 1)$ is thus simply the sum of the products of the pure gauge partition function and the matter partition function in each vacuum separately, 
\bel{
  Z_{L(p, 1)} = \sum_{\b N} Z\_{CS}(\b N) Z\_M(\b N).
}
It is now possible to analyze the gauge and the matter sectors separately.

\subsection{The gauge sector} 

The gauge sector in a particular vacuum can be reduced to a matrix model \cite{Aganagic:2002wv, Marino:2009dp, Marino:2011nm}, and its partition function can be expressed as the integral
\bel{\label{Z cs}
  Z\_{CS}(\b N) = \frac{e^{-\frac ig S\_{cl}(\b N) - \frac{ig}{12}N(N^2 - 1)}}{\prod_I N_I!}  \int  \frac{\d^N x}{(2\pi)^N} e^{\frac i{2g} (x - 2\pi i n(\b N)/p)^2} \prod_{1\leq i < j\leq N} \left(2 \sinh \frac{x_i - x_j}2\right)^2,
}
where $k$ is the CS level, $g = 2\pi/p(k + N)$ is the ``string coupling,'' $S\_{cl}(\b N) = (2\pi^2/p^2)\sum_I I^2 N_I$ is the classical CS action in the vacuum $\b N$, and $n(\b N)$ is an $N$-vector whose first $N_0$ entries are 0, the next $N_1$ entries are $1$, and so on. We are interested in the limit of zero 't Hooft coupling, $k \gg N$, where we only need the leading order behavior in the saddle point approximation. Upon analytic continuation to the complex plane, we may let $x = 2\pi i n(\b N)/p + \xi \sqrt g$ and take $g \approx 2\pi/kp \rar 0$, reducing the integral in \eqref{Z cs} to
\bel{
  \prod\_{off\ b.\;d.} \left(2i\sin \frac{\pi}p \left(n_i(\b N) - n_j(\b N)\right) \right)^2\int \d^N \xi \left(\frac{\sqrt g}{2\pi}\right)^N e^{i\xi^2/2} \prod\_{on\ b.\;d.} \left(\big(\xi_i - \xi_j\big)^2 g\right),
}
where ``on/off b.~d.'' refers to the pairs of numbers $(i, j)$ with $1\leq i < j \leq N$ that label elements of an $N\times N$ matrix that, respectively, do/do not fall into block-diagonal entries with $p$ blocks of dimensions $N_I \times N_I$. There are $(\b N^2-N)/2$ factors of $g$ in the product of ``on block diagonal'' entries, and hence the entire integral scales as $g^{\b N^2/2}$. The gauge sector thus has the free energy $-\ln Z\_{CS}(\b N)$ whose real part, the weight of the vacuum $\b N$ in the partition function, is given by
\algnl{\label{F cs}
  F\_{CS}(\b N) 
  &= \frac{\b N^2}{2} \ln \frac{kp}{2\pi} - \sum_{I < J} 2N_I N_J \ln\left\{ 2\sin\frac\pi p (J - I)\right\} - \sum_I \ln \left\{\prod_{n = 1}^{N_I} n!\right\},
}
where we have computed the integral over $\xi$'s by analytically continuing and transforming it into a Gaussian integral over $N\times N$ Hermitian matrices $\Xi$,
\bel{
  \frac1{N!}\int \frac{\d^N \xi}{(2\pi)^N} \ e^{-\xi^2/2}\! \prod_{1\leq i < j\leq N} (\xi_i - \xi_j)^2 = \frac1{\trm{Vol}(U(N))}\int \d \Xi\ e^{-\frac12\Tr\: \Xi^2} = \frac{\prod_{n = 1}^N n!}{(2\pi)^{N/2}}.
}
Note that we have dropped the irrelevant $\b N$-independent constants, like $(N/2)\ln\{2\pi\}$, from the free energy.

At infinite level the dominant term is $\b N^2 \ln k$. Minimizing it picks out the dominant vacuum $\b N_0 = (N/p, \ldots, N/p)$. This has been noted for $p = 2$ in \cite{Marino:2009dp}, where detailed saddle-point expansions have been examined for various complex values of $g$.  If $N/p$ is not an integer, i.e.~if $N = N' p + q$, the lowest-lying states consist of those vacua that have $q$ sectors with $N_I = N' + 1$ and $p - q$ sectors with $N_I = N'$. There are $\binom{p}{q}$ such vacua, and their energy difference is of order $N^0$. The matter sector will exhibit the same order of energy spacing in this almost-degenerate set of vacua. The total partition function at $k \gg N$ thus consists of a sum over the $\binom pq$ vacua that represent light excitations above the ground state, but for an explicit form we need to find the contribution from 
the matter sector as well.

\subsection{The matter sector}

The matter sector is represented by a free $U(N)$ model, i.e.~by a conformal theory of a complex $N$-vector field $\phi_a$. On a curved background, its Lagrangian is
\bel{
  \L = \phi_a^*\left(-\nabla^2 + \xi\mathcal R \right)\phi_a,
}
where $\nabla^2$ is the covariant Laplacian and $\xi \mathcal R$ is the conformal coupling to the curvature, needed to ensure the tracelessness of the stress tensor. On a lens space $L(p, 1)$ with $d = 3$, the curvature is constant and the conformal coupling equals
\bel{
  \xi \mathcal R = \frac{d - 2}{4(d - 1)} \cdot d(d - 1) = \frac34.
}
Thus, the matter sector in vacuum $\b N$ yields a contribution of $\ln\det\_{\b N}(-\nabla^2 + 3/4)$ to the total free energy $F = -\ln Z$. The subscript $\b N$ indicates that the conformal Laplacian acts on the space of $N$-vector fields whose $N_I$ components couple to the $U(N_I)$ connection. Each eigenstate of this operator is also an eigenstate of the conformal Laplacian on the three-sphere, and so we can calculate the needed functional determinant by summing over an appropriately restricted set of eigenvalues of the conformal Laplacian on $S^3$. This can be done as follows. 

The eigenvalues of the covariant Laplacian $\nabla^2$, given by $-\ell(\ell + 2) = 1 - (\ell + 1)^2$ for $\ell \geq 0$, each have degeneracy $(\ell + 1)^2$ on $S^3$. We wish to restrict ourselves to  the set of orbifold-invariant eigenstates. These are the states that acquire a trivial phase upon traversing this orbifolded circle of $S^3$.  If we let $m \in \{-\ell/2, -\ell/2 + 1, \ldots, \ell/2\}$ be the eigenvalue of the generator of rotations along the orbifolded circle, the eigenstates labeled by $m$ will acquire a phase of $e^{4\pi i m/p}$ upon traversing the circle. On the other hand, the existence of a nontrivial Wilson loop forces the eigenstates in the $I$-th sector of a given vacuum $\b N$ to acquire a Berry phase $e^{2\pi i I/p}$ by going around the same circle. Thus, the eigenstates of $\nabla^2$ on a lens space are those eigenstates of $\nabla^2$ on a sphere that  satisfy
\bel{\label{orb cond}
  e^{4\pi i m/p} e^{2\pi i I/p} = 1 \quad\trm{or}\quad 2m + I \in p\Z.
}
Each state whose quantum number $m$ satisfies this constraint will have a degeneracy of $\ell + 1$. The free energy of the matter sector in vacuum $\b N$ thus becomes
\algnl{\notag
  F\_M(\b N)
  &= \ln \det{\!}_{\b N} \left(-\nabla^2 + \frac34 \right)
  = \sum_I N_I \ln \det{\!}_I\left(-\nabla^2 + \frac34 \right) \\ \label{func det}
  &= \sum_I N_I \sum_{\ell = 0}^\infty (\ell + 1) d_I(\ell) \ln\left\{(\ell + 1)^2 - \frac14 \right\},
}
with $d_I(\ell)$ counting how many numbers $2m \in \{-\ell, -\ell + 2, \ldots, \ell\}$ satisfy the orbifold invariance condition \eqref{orb cond} by being congruent to $-I$ modulo $p$.

It is difficult to find a closed form for $d_I(\ell)$, but we can immediately note that $\sum_I d_I(\ell) = \ell + 1$. Therefore, if $p$ divides $N$, the CS-preferred vacuum $\b N_0 = (N/p, \ldots, N/p)$ has the matter free energy of the free $U(N/p)$ model on $S^3$ \cite{Marino:2011nm, Klebanov:2011gs}
\bel{
  F\_M(\b N_0) = \frac Np \sum_{\ell = 0}^\infty (\ell + 1)^2 \ln\left\{(\ell + 1)^2 - \frac14 \right\} = \frac{N}{8p} \left(\ln 4 - \frac{3\zeta(3)}{\pi^2}\right).
}
The case when $N/p$ is not an integer requires a bit more work. The free energy of a single scalar field in the $I$-th sector, denoted $F\_M(p, I) = \ln \det_I(-\nabla^2 + 3/4)$, can be computed by an appropriate reformulation and subsequent renormalization of the sum \eqref{func det}. This is done in the appendix, where it is found that $F\_M(p, I)$ can always be expressed in analytic form using polylogarithmic and $\zeta$-functions.

Once $F\_M(p, I)$ is known, the full matter free energy with $N = N' p + q$ can be written as
\bel{
  F\_M(\b N) = \frac {N'}{8} \left(\ln 4 - \frac{3\zeta(3)}{\pi^2}\right) + \sum_{I'} F\_M(p, I'),
}
where $I'$ runs over the $q$ sectors that have $N' + 1$ components. Using the CS free energy \eqref{F cs} we find that the total free energy is
\algnl{\notag
  F(\b N)
  &= \frac{p \left(N'\right)^2 + 2 q N' + q}{2} \ln \frac{kp}{2\pi} -  2\left(N'\right	)^2 \sum_{I < J} \ln\left\{ 2\sin\frac\pi p (J - I)\right\} -  p \ln \left\{\prod_{n = 1}^{N'} n!\right\} - \\ \notag
  &\quad -  N' \left[ q \sum_{I\neq 0} \ln\left\{ 2\sin\frac{\pi I}p \right\} + \frac {1}{8} \left(\ln 4 - \frac{3\zeta(3)}{\pi^2}\right) \right] - q\ln\left\{N' + 1\right\}\ -\\
  &\quad - 2\sum_{I' < J'}\ln\left\{ 2\sin\frac\pi p (J' - I')\right\} + \sum_{I'} F\_M(p, I').
}
This is the free energy of any vacuum $\b N$ with $N'$ components in $p - q$ sectors and with $N' + 1$ components in $q$ sectors. This formula is organized by powers of $N'$ and $k$, and we see that the vacua with minimal $\b N^2$ are distinguished only by the $O(1)$ term
\bel{
  \eps(\b q) = \sum_{I'} F\_M(p, I') - 2\sum_{I' < J'}\ln\left\{ 2\sin\frac\pi p (J' - I')\right\},
}
where $\b q$ is the $q$-vector whose components are labels $I'$ of sectors with a $U(N' + 1)$ gauge group. This vector is the only degree of freedom left in the system and it represents the ``topological states'' that are preserved even at infinite level. With $F_0 = F(\b N) - \eps(\b q)$, we may finally write the full partition function as 
\bel{\label{Z}
  Z_{L(p, 1)} = e^{-F_0} \sum_{\b q} e^{-\eps(\b q)}.
}

\section{Summary and implications}

Our main result is that the partition function of the singlet vector model on $L(p, 1)$ can be written as the sum \eqref{Z} over  light ($\eps \sim O(1)$) topological states with a Casimir energy $F_0 \sim \left(N'\right)^2 \ln k$. The number of such states, $\binom pq$, depends on number-theoretic relations between $N$ and $p$. It is of special note that the light states disappear when $q = 0$; this shows that the large $N$ limit must be taken with care, keeping track of the divisibility of $N$ by $p$ in order to retain knowledge of the correct number of light states.  Moreover, at $p > N$, we have $N' = 0$ and $q = N$, so there are $\binom p N$ topological states. At $p \gg N$ this number is $\sim p^{N - 1}$ and we find an exponential proliferation of light states --- independent of any number-theoretic conditions on $N$ and $p$. It should be emphasized that none of our results depend on taking the large $N$ limit. The topological states are an $O(N^0)$ phenomenon and should be taken into account 
as an $1/N$ effect in any large-$N$, zero-'t Hooft coupling calculation of gauge theories on topologically nontrivial spaces. In particular, such effects must be found in the gravity duals of large $N$ singlet vector models, believed to be described by 
Vasiliev higher spin theories. The absence of these light states in Vasiliev theories signifies that these gravity theories must be supplemented by new physics that we are yet to understand. It is conceivable that a stringy embedding of Vasiliev theory (based, perhaps, on the gravity dual of CS theory described in \cite{Gopakumar:1998ki}) would reveal that the gravity duals of extra light states are some analogs of fractional branes living on the orbifold singularity in the bulk, but it must be kept in mind that these are $O(1)$ effects, not $O(N)$ as one would expect for branes.

The singlet vector model that we have considered has no tunable parameters, and hence we cannot speak of phase transitions. However, one can imagine perturbing the conformal theory by various operators and computing the corresponding free energy. The ``band structure'' of the topological states will then be able to display nontrivial band crossing phenomena as we tune the deformations. However, ultimately the perturbations will take the theory to the $\phi^6$ fixed point or further to the Wilson-Fisher fixed point, and a new computation must be carried out in which one must regularize both the sums over eigenstates and the quartic coupling. We leave this for future work.

It is also possible to extend our results by moving away from zero 't Hooft coupling. An expansion in $1/k$ (or in $g \sim 1/(N + k)$) induces corrections in both the matter and the gauge sector. These $1/k$ gauge field-matter interactions allow for tunneling from the minimal energy vacua (labeled by $\b q$) to a new set of vacua in which $\b N^2$ is larger than before. These instanton corrections can, in principle, be computed through a straightforward expansion in Feynman diagrams on a sphere, generalizing the work of \cite{Aharony:2012nh}.

Finally, we comment on the bearing of this work on the dS/CFT correspondence developed in \cite{Anninos:2011ui, Anninos:2012qw, Anninos:2012ft}. The most operative version of this correspondence equates the amplitudes of $d = 3$, large $N$, anticommuting singlet vector model partition functions with the amplitudes of Hartle-Hawking wavefunctionals of four-dimensional de Sitter universes. The anticommutativity condition alters our discussion merely by changing $N \mapsto -N$ in the matter free energy. With this alteration, we may regard the $k \gg N \gg 1$ limit of the partition function $Z_{L(p, 1)}$ in \eqref{Z} as a relative probability amplitude for de Sitter universes having precisely the spatial geometry of $L(p, 1)$. This amplitude can be used to compare probabilities of spaces of different geometries, as proposed in \cite{Anninos:2012ft}, and in particular it could be used to compare probabilities of squashed vs.~non-squashed lens spaces. It is attractive to na\"ively extrapolate this procedure and 
use the full partition function $Z_{L(p, 1)}$ to compute relative probabilities between topologies. However, as shown in \cite{Maldacena:2011mk}, the gauge sector must be transformed by $N^2 \mapsto -N^2$ in order to get the Hartle-Hawking amplitude for the de Sitter dual. This is incompatible with the $N \mapsto - N$ transformation needed in the matter sector, and this suggests that the na\"ive extrapolation  fails for the coupled CS-matter sector. It would be interesting to formulate a consistent analytic continuation from the free energy of the $U(N)$ singlet vector model, dual to free energies of Vasiliev theories in asymptotically AdS spaces, onto the Hartle-Hawking amplitudes in asymptotically dS spaces. If the right prescription is discovered, our results will provide the wave functionals of asymptotically dS universes with a lens space topology.

\section*{Acknowledgments}

It is a pleasure to thank Dionysios Anninos, Shamik Banerjee, Daniel Harlow, Sean Hartnoll, Simeon Hellerman, Shamit Kachru, Raghu Mahajan, Shotaro Makisumi, Eva Silverstein, Douglas Stanford, and especially Stephen Shenker, for help and useful discussions.  The author is supported by the Stanford Institute for Theoretical Physics and an NSF Graduate Fellowship. 

\appendix
\section{Free energy of vector matter}

The purpose of this appendix is to explicitly calculate the free energy of the matter sector \eqref{func det},
\bel{\label{func det again}
  F\_M(p, \b N) = \ln \det{\!}_{\b N} \left(-\nabla^2 + \frac34 \right) = \sum_I N_I \ln \det{\!}_I\left(-\nabla^2 + \frac34 \right) \equiv \sum N_I F\_M(p, I),
}
where $\det\!_I$ is understood to contain only the states that satisfy the orbifold invariance condition \eqref{orb cond}. The following counting of such states makes the task tractable. Given any integer $i$ and a positive integer $j$, a state with $2m = -I + ip$ will appear precisely $\ell + 1$ times in each multiplet with $\ell = 2j + |ip - I|$. The functional determinant of the $I$-th sector, appearing in \eqref{func det again}, can then be written as
\algnl{\notag
  F\_M(p, I) &=
  \sum_{\substack{i \in \Z\\ j \geq 0}} \big(2j + |ip - I| + 1\big) \ln\left\{(2j + |ip - I| + 1)^2 - \frac14 \right\} \\ \notag
  &= \sum_{j \geq 0} (2j + I + 1)\ln\left\{(2j + I + 1)^2 - \frac14 \right\} + \\ \label{func det I}
  &\quad+ \sum_{\substack{i > 0\\ j \geq 0}}\left[ \big(2j + ip - I + 1\big) \ln\left\{(2j + ip - I + 1)^2 - \frac14 \right\} + (I \leftrightarrow -I)\right].
}
We may express this as a single infinite summation over  all possible values of $n(i, j) = 2j + ip$. We merely need to find the number of pairs $(i, j)$ that give the same sum $n(i, j)$. The number $n(i, j)$ can always be written as
\bel{\label{nij}
  n(i, j) = 2j + ip = np + J,
}
with $n$ being a positive integer (the case $n = 0$ is excluded because $i > 0$, and so $2j + ip \geq p$ at all times). To find the desired number of pairs $(i, j)$, we must now separately consider the cases when $p$ is even and when $p$ is odd.  

Let us first assume that $p$ is even. The equality \eqref{nij} forces $J$ to be even, and we may write $j = (J + (n - i)p)/2$. Therefore, for each $i \leq n$ there exists a positive $j$ such that \eqref{nij} is fulfilled. This means that there are precisely $n$ pairs of points $(i, j)$ for which $n(i, j)$ is equal to the given even number $np + J$.

Now assume that $p$ is odd. The number $(n - i)p + J$ still must be even. Thus, given $n$ and $J$, we will find one appropriate $j$ for each $i \leq n$ for which $n - i$ and $J$ have the same parity. The number of  pairs $(i, j)$ yielding the given $n(i, j) = np + J$ thus depends on the parity of $J$; it equals $\lceil n/2\rceil$ if $J$ is even and $\lfloor n/2 \rfloor$ if $J$ is odd. These results are summarized in Table \ref{T1}.

\begin{table}[bt] 
\begin{center}
\begin{tabular}{c|c} 
 {\bf $p$ even} & any $n$ \\ \hline
 $J$ even & $n$\\ \hline
 $J$ odd & 0 \\
\end{tabular}
\qquad
\begin{tabular}{c|c|c} 
 {\bf $p$ odd} & $n$ even & $n$ odd \\ \hline
 $J$ even & $n/2$ & $(n + 1)/2$\\ \hline
 $J$ odd & $n/2$ & $(n - 1)/2$ \\
\end{tabular}
\caption{Number of pairs $(i, j)$ giving the same combination $2j + ip = np + J$ for a given $n$.}
\label{T1}
\end{center}
\end{table}

It is thus possible to creatively split the sum over possible values of $n(i, j)$ so as to get only one infinite sum. If $p$ is even, we may write
\bel{
  \sum_{\substack{i > 0\\ j \geq 0}} f(2j + ip) = \sum_{n = 1}^{\infty} \sum_{\trm{even}\;J} n f(np + J). 
}
If $p$ is odd, on the other hand, we have
\algnl{\notag
  \sum_{\substack{i > 0\\ j \geq 0}} f(2j + ip) 
  &= \sum_{n = 1}^{\infty} \sum_{J} n f(2np + J) + \\ 
  &\quad + \sum_{n = 1}^{\infty}\left[\sum_{\trm{even}\;J} n f\big((2n - 1)p + J\big) + \sum_{\trm{odd}\;J} (n - 1) f\big((2n - 1)p + J\big) \right].
}
In particular, if $p$ is even, the matter free energy \eqref{func det I} can be written as
\algnl{\notag
  F\_M(p, I)
  &= \sum_{n = 0}^\infty (2n + I + 1)\ln\left\{(2n + I + 1)^2 - \frac14 \right\} + \\ \label{F even p}
  &\quad+ \sum_{\trm{even}\;J} \sum_{n = 1}^\infty\left[ n\big(np + J - I + 1\big) \ln\left\{(np + J - I + 1)^2 - \frac14 \right\} + (I \leftrightarrow -I)\right],
}
and if $p$ is odd we can write
\algnl{\notag
  F\_M(p, I) 
  &= \sum_{n = 0}^\infty (2n + I + 1)\ln\left\{(2n + I + 1)^2 - \frac14 \right\} + \\ \notag
  &\hspace{-1em}+ \sum_{J} \sum_{n = 1}^\infty\left[ n\big(2np + J - I + 1\big) \ln\left\{(2np + J - I + 1)^2 - \frac14 \right\} + (I \leftrightarrow -I)\right] + \\ \notag
  &\hspace{-1em}+ \sum_{J} \sum_{n = 1}^\infty\left[ n\big(2np - p + J - I + 1\big) \ln\left\{(2np  - p + J - I + 1)^2 - \frac14 \right\} + (I \leftrightarrow -I)\right] - \\ \label{F odd p}
  &\hspace{-1em}- \sum_{\trm{odd}\;J} \sum_{n = 1}^\infty\left[\big(2np - p + J - I + 1\big) \ln\left\{(2np  - p + J - I + 1)^2 - \frac14 \right\} + (I \leftrightarrow -I)\right].
}

All of these sums are divergent and need to be renormalized. We can use $\zeta$-regularization to find the finite value of the sums
\bel{\label{sums}
  \sum_{n = 0}^\infty n (np + \alpha) \ln\left\{(np + \alpha)^2 - \frac14\right\} \quad\trm{and}\quad \sum_{n = 0}^\infty (np + \alpha) \ln\left\{(np + \alpha)^2 - \frac14\right\};
}
their linear combinations will yield all the sums appearing in the free energy. The procedure is standard \cite{Marino:2011nm}. If we define
\bel{\label{zetas}
  \zeta_1(s; p, \alpha) = \sum_{n = 1}^\infty \frac{n (np + \alpha)}{\big((np + \alpha)^2 - 1/4\big)^s} \quad\trm{and}\quad \zeta_2(s; p, \alpha) = \sum_{n = 1}^\infty \frac{np + \alpha}{\big((np + \alpha)^2 - 1/4\big)^s},
}
the sums \eqref{sums} will be given by $-\zeta'_{1/2}(0; p, \alpha)$. For instance, the free energy of the matter sector at even $p$, given by \eqref{F even p}, will be
\algnl{\notag
  F\_M(p, I)
  &= (I + 1)\ln\left\{(I + 1)^2 - \frac14 \right\} - \\ 
  &\quad - \der{}s\biggr|_{s = 0} \left(\zeta_2(s; 2, I + 1) + \sum_{\trm{even}\;J} \big(\zeta_1(s; p, J + I + 1) + \zeta_1(s; p, J - I + 1)\big) \right). 
}
The odd $p$ case \eqref{F odd p} is treated the same way. The derivatives of the $\zeta$-sums may be explicitly calculated for any given $p$ and $I$ by splitting the sum into a convergent piece (where $\der{}{s}|_{s = 0}$ can be commuted through the sum, making it computable in practice) and a divergent piece expressed in terms of $\zeta$-functions of negative integers (which can be assigned unique finite values through analytic continuation). All the results can be expressed in analytic form. For instance, when $p = 2$ we find the free energies
\bel{
  F\_M(2, 0) = -\frac{C}{\pi} + \frac1{16}\left(\ln 4 - \frac{3\zeta(3)}{\pi^2}\right),\quad F\_M(2, 1) = \frac{C}{\pi} + \frac1{16}\left(\ln 4 - \frac{3\zeta(3)}{\pi^2}\right),
}
where $C = \sum_{n = 0}^\infty (-1)^n/(2n + 1)^2 \approx 0.916$ is Catalan's constant. The free energies at higher $p$ can similarly be expressed in terms of polylogarithms. There seems to exist no convenient closed form for $F\_M(p, I)$ at arbitrary $p$ and $I$, and hence, as customary, the numerical values of these results are tabulated and can be found in Table \ref{T2}. As a nontrivial check of our renormalization scheme, note that summing the free energies $F\_M(p, I)$ over all $I = 0, 1, \ldots, p - 1$ and for any $p$ gives $\frac18\left(\ln 4 - 3\zeta(3)/\pi^2 \right) \approx 0.128$, the free energy of a complex scalar field on a sphere \cite{Marino:2011nm, Klebanov:2011gs}. Using these data it is possible to explicitly determine the ground state vacuum that minimizes the total free energy of the theory.

\begin{table}
\begin{small}
\begin{tabular}{c|ccccccccc}
 $p\backslash I$ & 0 & 1 & 2 & 3 & 4 & 5 & 6 & 7  \\ \hline \\
 1& 0.127614 \\
 2& -0.227754 & 0.355368 \\
 3& -0.511577 & 0.319596 & 0.319596 \\
 4& -0.813519 & 0.177684& 0.585765& 0.177684\\
 5& -1.15495& -0.0446098& 0.685894& 0.685894& -0.0446098 \\
 6& -1.54434& -0.338697& 0.658292& 1.03276& 0.658292& -0.338697 \\
 7& -1.9859& -0.700683& 0.523212& 1.23423& 1.23423 &
   0.523212& -0.700683 \\
 8& -2.48208& -1.12847& 0.292883& 1.30615& 1.66856& 1.30615& 
    0.292883& -1.12847  \\
\end{tabular}
\end{small}
\caption{Values of the matter free energy $F\_M(p, I)$ in different sectors and on different lens spaces.}
\label{T2}
\end{table}

The $\zeta$-regularization procedure obfuscates the $I$-dependence of the free energy $F\_M(p, I)$. Heat kernel regularization may be used to write the matter free energy as a transparent effective action for the Wilson loop $e^{2\pi i I/p}$ along the orbifolded circle \cite{Banerjee:2012gh}. It might be useful to think in terms of such effective actions, so here we outline this alternative method of integrating out matter.  Letting $\b x$ be the coordinates on $S^3$, we may write the free energy of a scalar field as the one loop integral
\bel{
  F\_M(p, I) = \ln \det\!_I \left(-\nabla^2 + \frac34 \right) = - \int_0^\infty \frac{\d s}{s} \int_{S^3}\d^3 \b x\ G_I(\b x, s; \b x, 0), 
}
where 
\bel{
  G_I(\b y, s; \b x, 0) = \sum_J V_I\left(\b y + \frac{4\pi J} p\b e_\psi; \b x\right)\ \left\langle\b y + \frac{4\pi J}p\b e_\psi\ \biggr|\ e^{-s(-\nabla^2 + 3/4)}\ \biggr|\ \b x\right\rangle
}
is the scalar field propagator in the $I$-th sector of a given vacuum on $L(p, 1)$, lifted to the three-sphere and summed over all points that are identified by the orbifold operation; $V(\b y; \b x)$ is the $I$-th sector eigenvalue of the Wilson line along any path connecting $\b x$ and $\b y$, and $\qmat{\b y}{\cdot}{\b x}$ is the propagator on the sphere that can be evaluated by expanding the matrix element in terms of spherical harmonics. For the case at hand, we find 
\bel{
  F\_M(p, I) =  \sum_J \left(-	\int_0^\infty \frac{\d s}s \sum_{\ell = 0}^\infty (\ell + 1) e^{-s(\ell(\ell + 2) + 3/4)} \sum_{m = -\ell/2}^{\ell/2} e^{4\pi i m J/p}\right) e^{2\pi i I J/p}.
}
The coefficients in this ``Fourier expansion'' can now be calculated by renormalizing the infinite sum and then integrating over the propagation times $s$.

\end{document}